# From ($\pi$, 0) magnetic order to superconductivity with ($\pi$, $\pi$) magnetic resonance in $Fe_{1.02}(Te_{1-x}Se_x)$

T.J. Liu[1], J. Hu[1], B. Qian[1], D. Fobes[1], Z.Q. Mao[1*], W. Bao[2], M. Reehuis[3], S.A.J. Kimber[3], K. Prokes[3], S. Matas[3], D.N. Argyriou[3], A. Hiess[4], A. Rotaru[5], H. Pham[5], L. Spinu[5], Y. Qiu[6,7], V. Thampy[8], A.T. Savici[8], J. A. Rodriguez[8], and C. Broholm[6,8]

[1] Department of Physics and Engineering Physics, Tulane University, New Orleans, Louisiana 70118, USA

[2] Department of Physics, Renmin University of China, Beijing 100872, China

[3] Helmholtz-Zentrum Berlin für Materialen und Energie, Hahn-Meitner Platz 1, D-14109 Berlin, Germany

[4] Institut Max von Laue-Paul Langevin, 6 rue Jules Horowitz, BP 156, F-38042, Grenoble Cedex 9, France

[5] Advanced Materials Research Institute and Department of Physics, University of New Orleans, New Orleans, Louisiana 70148, USA

[6] NIST Center for Neutron Research, National Institute of Standards and Technology, Gaithersburg, Maryland 20899, USA

[7] Department of Materials Science and Engineering, University of Maryland, College Park, Maryland 20899, USA

[8] Institute for Quantum Matter and Department of Physics and Astronomy, The Johns Hopkins University, Baltimore, Maryland 21218 USA

The iron chalcogenide $Fe_{1+y}(Te_{1-x}Se_x)$ is structurally the simplest of the Fe-based superconductors[1-3]. Although the Fermi surface is similar to iron pnictides[4-5], the parent compound $Fe_{1+y}Te$ exhibits antiferromagnetic order with in-plane magnetic wave-vector ($\pi$, 0)[6]. This contrasts the pnictide parent compounds where the magnetic order has an **in-plane magnetic wave-vector ($\pi$, $\pi$) that connects hole and electron parts of the Fermi surface[7-8]. Despite these differences, both the pnictide and chalcogenide Fe-superconductors exhibit superconducting spin resonances around ($\pi$, $\pi$)[9-11], suggesting a common symmetry for their superconducting order parameter. A central question in this burgeoning field is therefore how ($\pi$, $\pi$) superconductivity can emerge from a ($\pi$, 0) magnetic instability [12]. Here, we report that the magnetic soft mode evolving from the ($\pi$, 0)-type magnetic long-range order is associated with weak charge carrier localization. Bulk superconductivity occurs only as the magnetic mode at ($\pi$, $\pi$) becomes dominant upon doping. Our results suggest a common magnetic origin for superconductivity in iron chalcogenide and pnictide superconductors.**

Unconventional superconductivity in cuprates, heavy fermion intermetallics, and strontium ruthenate all appear in close proximity to magnetic instabilities and appears to be mediated by spin fluctuations. The newly discovered iron pnictide superconductors[13-16] follow the paradigm of superconductivity achieved by suppressing a long-range magnetic order through charge carrier doping or pressure. The long-range antiferromagnetic (AFM) order in the parent compounds of iron pnictide superconductors is characterized by the in-plane Fermi surface nesting wave-vector $Q_n = (\pi, \pi)$[7-8]. (Here and throughout the paper we refer to wave vectors in units of the inverse tetragonal lattice parameters.) Iron chalcogenide $Fe_{1+y}(Te_{1-x}Se_x)$ superconductors, discovered more recently[1-3], have a similar Fermi surface as that of iron pnictides, according to both density functional calculations[4] and photoemission measurements [5]. However, the undoped parent compound of this system, $Fe_{1.02}Te$, exhibits an AFM order characterized by an in-plane wave vector $Q_m = (\pi, 0)$[6], which distinguishes this compound from the iron pnictide parent materials. Yet both doped iron chacolgenide [11] and iron pnictide [9-10] superconductors exhibit a magnetic resonance in the spin excitation spectra below $T_c$ around the wave vector $(\pi, \pi)$, consistent with $s_\pm$ pairing symmetry [17-19]. Resolution of the dichotomy between $(\pi, 0)$ magnetic order in undoped FeTe and superconductivity with $(\pi, \pi)$ magnetic resonance in Se doped samples is a key challenge to our emerging understanding of iron based superconductivity [12].

In this work, we address the challenge through systematic investigation of transport, magnetic, and superconducting properties in various regions of the phase diagram of $Fe_{1.02}(Te_{1-x}Se_x)$ via resistivity, Hall coefficient, magnetic susceptibility, specific heat, and neutron scattering measurements. We find that magnetic correlation of the $(\pi, 0)$ variety survive as short-range magnetic fluctuations after the long-range ordered AFM phase has been suppressed by partial Se substitution for Te. These correlations appear to suppress bulk superconductivity and lead to weak charge carrier localization in under-doped sample. Bulk superconductivity occurs only when magnetic correlations near $(\pi, 0)$, though still present, are strongly suppressed and spin fluctuations near $(\pi, \pi)$ become dominant for $x > 0.29$, with the latter spin fluctuations exhibiting a spin gap and a spin resonance in the superconducting state [11,20]. These results indicate that short-range magnetic fluctuations near $(\pi, 0)$ are incompatible with superconductivity and that the superconducting mechanisms of iron chalcogenides and iron pnictides have similar origins associated with $(\pi, \pi)$ spin-fluctuations.

Using a variety of techniques we have constructed a detailed electronic and magnetic phase diagram of $Fe_{1.02}(Te_{1-x}Se_x)$ with $0 \leq x < 0.5$, which is shown in Figure 1**a** . In summary, we find three composition regions with distinct physical properties. Region I ($0 \leq x < 0.09$) exhibits long range AFM order with wave vector $(\pi, 0)$, while Region II ($0.09 < x < 0.29$) displays short range magnetic correlations at the same wave vector. Samples from both these regions exhibit non-bulk superconductivity. *Only* in Region III ($x \geq 0.29$) do we find evidence of bulk superconductivity. More specifically in Fig. 1**a**, symbols ■, ▼, ▲, +, represent the Néel temperatures $T_N$ determined by neutron diffraction, dc susceptibility, Hall coefficient, and resistivity measurements respectively. Importantly, these disparate measurements are entirely consistent with each other. $T_N$ gradually decreases with increasing Se content, approaching zero as $x$ is increased to ~ 0.09. A trace of superconductivity is observed for $0.04 \leq x < 0.09$, which will be examined in greater detail later in Fig. 3. The symbol ◊ represents the onset of superconducting transition $T_c^\rho$ probed by resistivity. In Region II ($0.09 < x < 0.29$), though long-range AFM order is fully suppressed, superconductivity remains non-bulk and the superconducting volume fraction, $V_{SC}$, is less than 3% for all samples with $x < 0.29$. In Region III ($x \geq 0.29$), however, bulk superconductivity is found. Symbol ♦ represents the bulk superconducting transition temperature $T_c^\chi$ probed by susceptibility. $V_{SC}$ rises to above 75 % for $x \geq 0.29$, as shown in Fig. 1**b**. In Region II, the transport properties above $T_c^\rho$ indicate weak charge localization which contrasts with the metallic behavior seen in the normal state of Region III and the AFM phase of Region I. The cross over is clearly indicated by the sign change in the derivative of resistivity with respect to temperature $d\rho/dT$ (see Fig. 1**b**). The absence of bulk superconductivity in Region II makes the phase diagram of $Fe_{1.02}(Te_{1-x}Se_x)$ distinct from those of iron pnictide superconductors where bulk superconductivity either appears immediately following suppression of long-range AFM order [21-22], or putatively coexist with ($\pi$, $\pi$) AFM order in a certain composition range [23-25].

We shall now address properties of each region of the phase diagram (Fig. 1**a**) in greater detail. Throughout the AFM phase (Region I), neutron scattering measurements reveal the same commensurate ($\pi$, 0) magnetic structure as reported for the parent compound [6]. The ordered magnetic moment of iron $M_{Fe}$ depends strongly on Se content as shown in Fig. 2**a**. $M_{Fe}$ is approximately 2.09(3) $\mu_B$/Fe for the $x = 0.04$ sample, but decreases to 0.33(2) $\mu_B$/Fe for the $x = 0.08$ sample. The saturated staggered magnetic moment for $Fe_{1.02}Te$ is much larger than that of iron pnictide parent compounds (for example, 0.36 $\mu_B$/Fe for LaOFeAs[7] and 0.87 $\mu_B$/Fe for $BaFe_2As_2$[8]). This implies that the mechanism for magnetism in $Fe_{1+y}Te$ is different from that of pnictide parent materials [26-27]. Hall effect measurements shown in Fig. 2**b** indicate that the AFM transition in Region I is accompanied by a remarkable change of the Fermi surface. For $x < 0.08$ the Hall coefficient $R_H$ exhibits a sharp drop from a positive to a negative value across the transition. This indicates that the Fermi surface is dominated by holes above $T_N$ and by electrons below $T_N$.

Figure 3**a** presents the in-plane resistivity $\rho_{ab}(T)$ as a function of temperature for typical samples in Region I. $\rho_{ab}(T)$ exhibits an anomaly at $T_N$, which is marked by a downward arrow in the figure. Each sample in this region also shows a trace of superconductivity below the AFM transition. This can be seen from the second drop of $\rho_{ab}(T)$ at low temperatures (denoted by upward arrows in Fig. 3**a**). While $T_c^\rho$ shows a systematic increase with increasing Se content in this region, the superconducting volume fraction is nearly zero (Fig. 1**b** and 3**d**) since we did not observe any diamagnetism below $T_c^\rho$ for these samples (see Fig. 3**d**). The absence of bulk superconductivity in Region I is also demonstrated by the neutron scattering data shown in Fig. 2**a**. We do not observe any drop in the ordered magnetic moment below $T_c^\rho$. This is opposite to the behavior seen in Co-doped $BaFe_2As_2$ where the magnetic order parameter exhibits a sharp decrease below $T_c$ in the region of coexistence between magnetism and superconductivity [25].

Despite the complete suppression of long range AFM order, superconductivity remains a non-bulk phenomenon throughout Region II. Although all samples in this region exhibit zero resistance below $T_c^\rho$, their susceptibility fails to display significant diamagnetism when the resistivity vanishes (Fig. 1**b** and 3**d**). The superconducting volume fraction of these samples estimated from $-4\pi\chi$ is thus below 3%. Further, the specific heat of samples in both Regions I and II are free of anomalies near the resistive superconducting transition. They can approximately be described by $C = \gamma T + \beta T^3$ at low temperatures, where $\gamma T$ and $\beta T^3$ represent the electron and phonon specific heat respectively. Figure 3**e** shows data for typical samples. The electronic specific coefficient $\gamma$ derived from linear fitting for various samples is given in the left inset to Fig. 3**e**. The right inset to Fig. 3**e** displays an example of the fit for the $x = 0.19$ sample where we observe a slight deviation from linearity below 4.5 K which may be due to non-bulk superconductivity. The significant increase of $\gamma$ for $x \geq 0.09$ is associated with enhanced magnetic fluctuations as shown below.

In contrast, samples in Region III exhibit characteristics of bulk superconductivity. Their susceptibility exhibits significant diamagnetism below $T_c^\chi$ and anomalous peaks near $T_c^\chi$ in specific heat data also indicates a bulk phase transitions (see Fig. 3**d** and 3**e**). The inferred superconducting volume fraction rises to above 75% for $x \geq 0.29$ (Fig. 1**b**). In addition, samples with bulk superconductivity in Region III differ from samples with non-bulk superconductivity in Region II in their normal state properties as noted above. As shown in Fig. 3**b** and 3**c**, the normal state in Region III exhibits metallic behavior in $\rho_{ab}(T)$. However, samples in Region II display a noticeable non-metallic upturn prior to the superconducting transitions in $\rho_{ab}(T)$. When $T_c < T < 20$ K, $\rho_{ab}(T)$ is characterized by a logarithmic temperature dependence ( see supplemental information), indicating weak charge carrier localized in Region II.

Why is bulk superconductivity suppressed and charge carriers weakly localized in Region II? It is a critical question to understand the difference between iron chalcogenide and iron pnictide superconductors, so we examine several possible explanations in the following. One possibility is that the quenched disorder induced charge localization suppresses superconductivity. Less disorder in Region III than Region II would however be most surprising in this alloy system. We do however have evidence that the magnetic correlations are changing profoundly from region II to region III. Our early neutron scattering measurements revealed that when the long-range AFM order is suppressed, strong short-range fluctuating remain near ($\pi$, 0) [6], a behavior subsequently confirmed elsewhere[28-29]. To clarify the role of such short-range magnetic fluctuations, we performed neutron scattering measurements on two typical single crystal samples with $x = 0.19$ and 0.38. The $x = 0.19$ sample resides in Region II and has a superconducting volume fraction of ~ 2%, whereas the $x = 0.38$ sample is in Region III and has a superconducting volume fraction of ~90%. As shown in Fig. 4**a** and 4**b**, both samples exhibit quasi-elastic scattering centered near ($\pi$, 0). With the data normalized to phonon intensity, ($\pi$, 0) magnetism in the $x = 0.19$ sample is significantly stronger than in the $x = 0.38$ sample, indicating that quasi-static short-range magnetic order of the ($\pi$, 0) variety is associated with region II and is suppressed in Region III. This short range order co-exists with ($\pi$, $\pi$) spin fluctuations for which a spin-gap and a magnetic resonance forms for the $x = 0.38$ sample in the bulk superconducting state. For a sample with $x = 0.27$, well into Region II Lumsden *et al*. also find spin excitations at the same wave vector [30], suggesting that magnetic scattering at ($\pi$, $\pi$) and ($\pi$, 0) co-exist over a wide range in composition.

To explore this co-existence, we performed inelastic neutron scattering measurements on an $x = 0.05$ single crystal. This sample exhibits long range ($\pi$, 0) magnetic order with an ordered moment of 1.68(6) $\mu_B$/Fe. In Fig. 4(c) we illustrate typical transverse inelastic neutron scattering scans centered at ($\pi$, $\pi$) which were measured up to 8 meV, and show the presence of well defined magnetic excitations similar to those we reported earlier for an $x = 0.4$ sample [20]. For the lower energy excitation at 2 meV we find a peak with a width broader than the $Q$-resolution of the instrument. The flat-top structure suggests that it consist of two components that separate through dispersion at higher energies. As found previously for an $x = 0.4$ sample, the corresponding dispersion relation extrapolates to incommensurate points ($1/2+\varepsilon$, $1/2-\varepsilon$) with $\varepsilon = 0.10(1)$, however, with a softer dispersion velocity of 62(5) meV Å as compared to 345(2) meV Å for the $x = 0.4$ bulk superconducting sample [20].

These results indicate that magnetic correlations characterized by $\boldsymbol{Q}_m = (\pi, 0)$ are not favorable to superconducting pairing on the contrary these quasi-static and low energy fluctuations may instead be pair breaking. This scenario is corroborated by our neutron scattering measurements performed on a sample $Fe_{1.11}(Te_{0.62}Se_{0.38})$, with excess Fe. Though the 38% Se content would place the sample in Region III, we found similar properties as for samples in Region II: Both susceptibility and specific heat measurements indicate that bulk superconductivity is suppressed; the resistivity follows a logarithmic temperature dependence indicating weakly localized charge carriers (see supplemental information), consistent with our previous report[31]. Our neutron scattering measurements (shown in the supplemental information) show an absence of low energy magnetic scattering at ($\pi, \pi$) but clearly defined magnetic short range ordering at ($\pi, 0$). This result indicates that magnetic fluctuations near ($\pi, 0$) are incompatible with superconductivity and instead are associated with weak charge carrier localization.

We have explored the dichotomy between ($\pi, 0$) and ($\pi, \pi$) magnetism in the $Fe_{1.02}(Te_{1-x}Se_x)$ system. For low Se content long range magnetic order is formed with a magnetic wave vector ($\pi, 0$). Dynamic magnetic correlations with a ($\pi, \pi$) wave vector however, do co-exist in the material. Increasing Se doping tunes the relative strength of these distinct correlations. However, as long range magnetic order at ($\pi$, 0) is replaced by short range quasi-static correlations, long range magnetic order at ($\pi$, $\pi$) is never observed in the system. It may be suppressed by increasing Se content as for the charge doped iron pnictides. For bulk superconducting samples it is at the ($\pi$, $\pi$) magnetic wave vector that a spin gap and a magnetic resonance is formed, a result that can be taken as confirmation of $s_\pm$ type superconductivity [17-19]. This indicates that both iron chalcogenide and iron pnictide superconductors have a similar magnetic mechanism for superconducting pairing. Our early work shows that the superconductivity of the $Fe_{1+y}(Te_{1-x}Se_x)$ system also depends on excess Fe content; increasing excess Fe content suppresses superconductivity [31], and therefore the phase boundary between non-bulk and bulk superconductivity in Fig. 1**a** depends on the excess Fe content $y$. The current boundary of $x = 0.29$ is valid for $y = 0.02$. If $y$ is larger than 0.1, the bulk superconductivity is suppressed even for $x > 0.29$ [31]. Overall, our phase diagram in Fig. 1**a** is consistent with previous reports of bulk superconductivity in $Fe_{1+y}(Te_{1-x}Se_x)$ single crystals [29,31-32].

In conclusion magnetic correlations near ($\pi$, 0) are not favorable to superconductivity but are associated with weak charge carrier localization in an intermediate region between long-range AFM and superconductivity. Bulk superconductivity occurs only in a composition region where ($\pi$, 0) magnetic correlation are strongly suppressed and ($\pi$, $\pi$) spin fluctuations associated with the nearly nesting Fermi surface dominate. This indicates that iron chalcogenide and iron pnictide superconductors, despite a competing magnetic instability in the former, have a similar mechanism for superconductivity.

*e-mail: zmao@tulane.edu

**Acknowledgements**

The work at Tulane is supported by the NSF under grant DMR-0645305 for materials and equipment, the DOE under DE-FG02-07ER46358 for personnel. Work at AMRI was supported by DARPA through Grant HR 0011-09-1-0047. Work at NIST is in part supported by the NSF under Grant DMR-0454672. Work at the Johns Hopkins University Institute for Quantum Matter is supported by the DOE under DE-FG02-08ER46544. DNA and KP, thanks the Deutsche Forschungsgemeinschaft for support under SPP 1458.

**Methods**

$Fe_{1.02}(Te_{1-x}Se_x)$ single crystals used in this study were synthesized using a flux method [31] and were shown to be tetragonal phase with the space group *P4/nmm* at room temperature by x-ray and neutron diffraction measurements [6]. The compositions of crystals were determined using energy dispersive x-ray spectrometer (EDXS). Since our early studies revealed that properties of this system are sensitive to Fe nonstoichiometry [6,31], we have chosen samples with the excess Fe less than 3% for all compositions to avoid complications caused by excess Fe. We measured resistivity with a four-probe method, Hall effect with a five-probe method, and specific heat with an adiabatic relaxation technique using commercial Physical Property Measurement System. DC magnetic susceptibility was measured using commercial SQUID. Neutron diffraction experiments were carried out on the 4-circle diffractometer E5 and the 2-axis diffractometer E4 at the BER II reactor at the Helmholtz-Zentrum, and neutron scattering measurements probing magnetic short-range order and excitations were performed using MACS and DCS instruments at NIST. Inelastic neutron scattering measurements were carried out on the $x = 0.05$ sample using the triple axis spectrometer IN8 operated by the Institut Laue-Langevin (ILL), with the sample mounted with the *c*-axis vertical giving us access to spin excitations within the basal *ab*-plane.

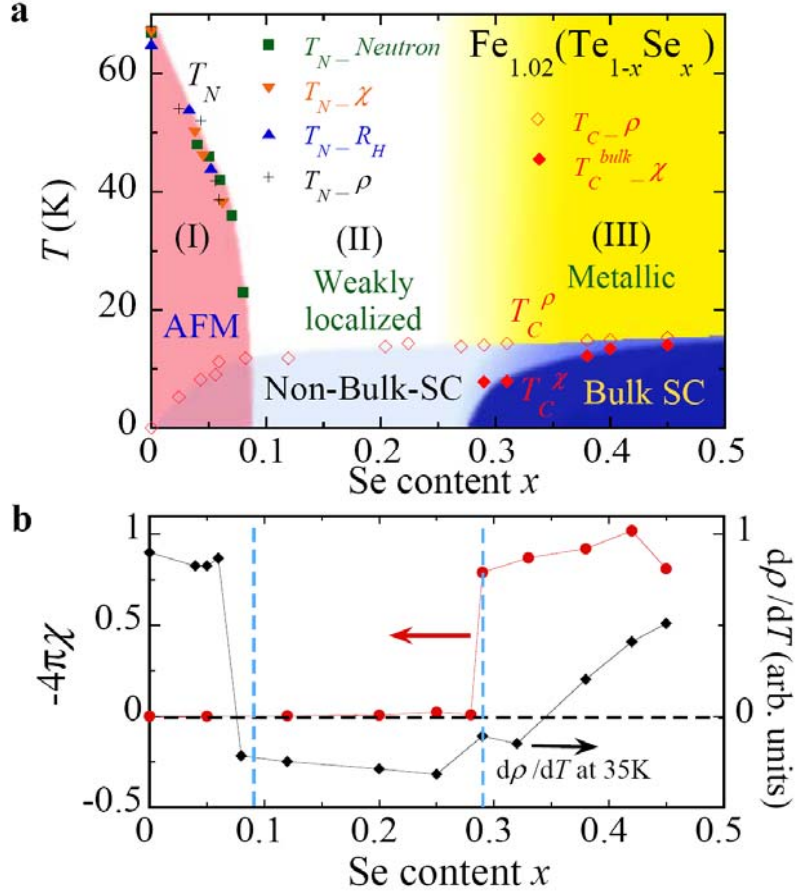

**Figure 1: Magnetic and superconducting properties of $Fe_{1.02}(Te_{1-x}Se_x)$ ($0 \leq x < 0.5$).**
**a.** The phase diagram. The Néel temperature, $T_N$, of antiferromagnetic phase (AFM), determined by neutron scattering (green squares), susceptibility (orange triangles), Hall coefficient (blue triangle), and resistivity (black crosses) measurements. $T_c^\rho$, onset of superconducting transition probed by resistivity (◊); $T_c^\chi$, bulk superconducting transition temperature (♦) probed by susceptibility. Bulk superconductivity (SC) exists when sufficient Te is replaced by Se, with the superconducting volume fraction > 75% for $x \geq 0.29$. For $x<0.29$, only non-bulk-SC exists with the superconducting volume fraction < 3 %. The bulk SC and non-bulk SC concentration regions also differ in their normal state transport property: metallic in the former, non-metallic in the latter. **b.** The superconducting volume fraction ($-4\pi\chi$) and the derivative of normalized resistivity ($\rho(T)/\rho(300K)$) with respect to temperature as a function of Se content.

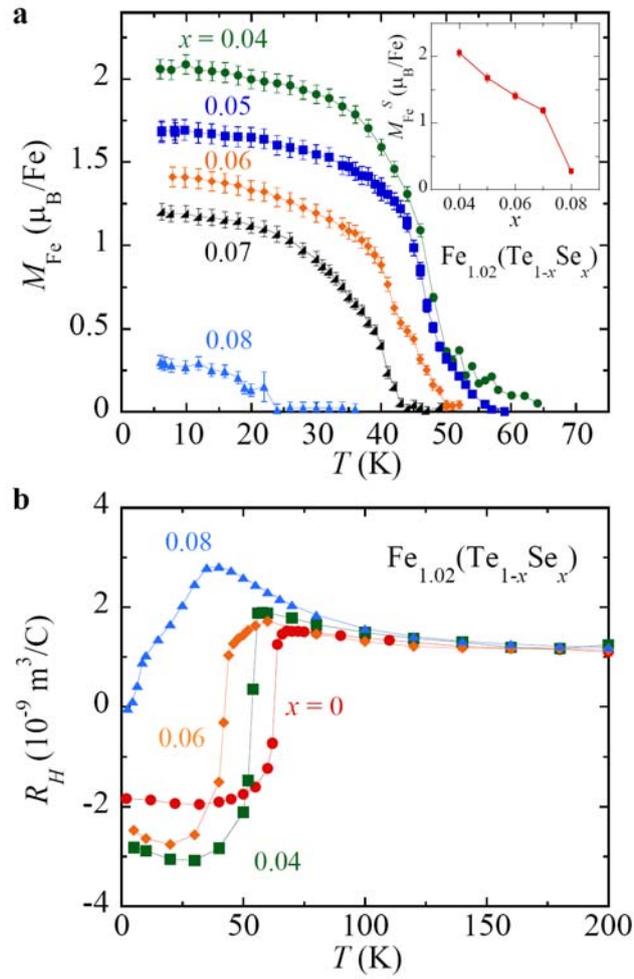

**Figure 2: Evolution of the long-range AFM order and FS variation across the AFM transition in Fe$_{1.02}$(Te$_{1-x}$Se$_x$). a**, Temperature dependence of the ordered magnetic moment $M_{Fe}$. Inset, the saturated moment M$_{Fe}$ as a function of Se content. The magnetic order is suppressed when $x > 0.09$. **b**, Hall coefficients as a function of temperature . The FS changes significantly across the AFM transition.

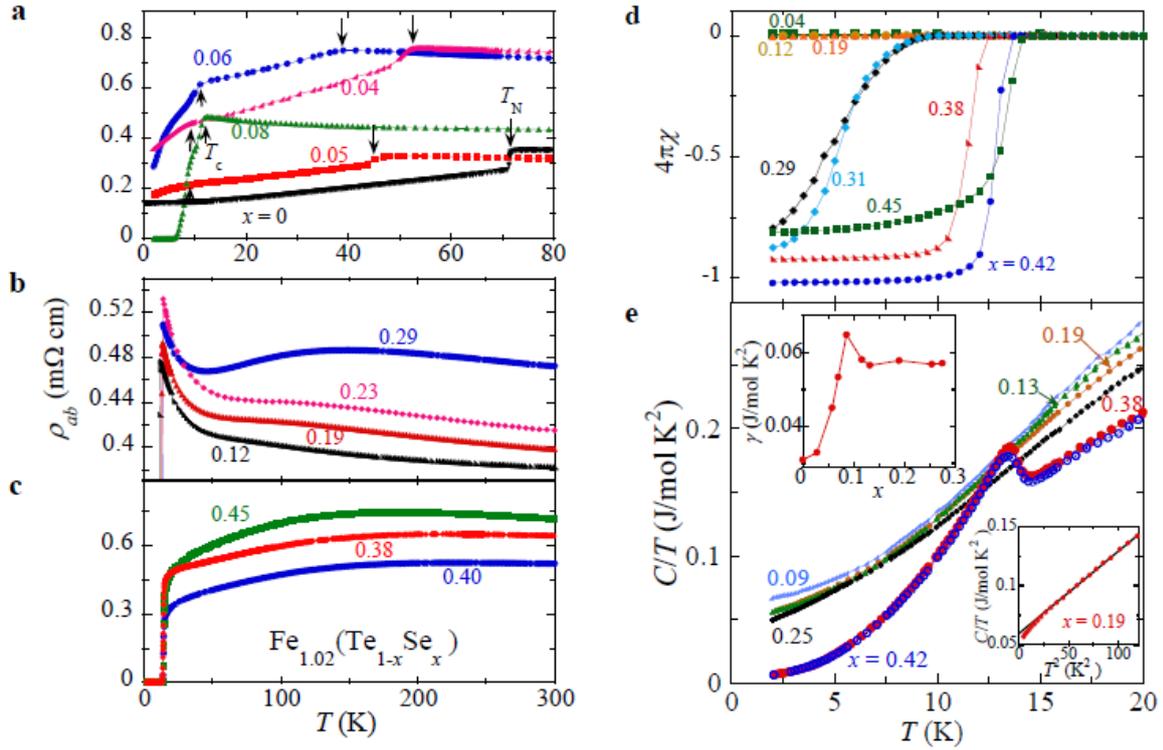

**Figure 3: Evolution of superconductivity as a function of Se content for $Fe_{1.02}(Te_{1-x}Se_x)$.**
**a**, In-plane resistivity $\rho_{ab}(T)$ as a function of temperature for samples in the AFM region ($0 \leq x < 0.09$). The downward arrows mark the AFM transition and the upward arrows mark the onset of a trace of superconductivity. **b**, $\rho_{ab}(T)$ for samples with $0.09 < x \leq 0.29$. **c**, $\rho_{ab}(T)$ for samples with $x > 0.29$. **d**, Magnetic susceptibility data measured with a zero-field-cooling history and a field of 30 Oe for typical samples. **e**, Specific heat divided by temperature $C/T$ as a function of temperature for various samples. The left inset is the electronic specific heat coefficient $\gamma$ as a function of Se content $x$. Right inset is $C/T$ as a function of $T^2$ for the $x = 0.19$ sample. Both magnetic susceptibility and specific heat data show that bulk superconductivity occurs only in samples with $x \geq 0.29$. The samples with bulk superconductivity exhibit metallic temperature dependence in $\rho_{ab}$, while those samples without bulk superconductivity display the non-metallic temperature dependence in $\rho_{ab}$.

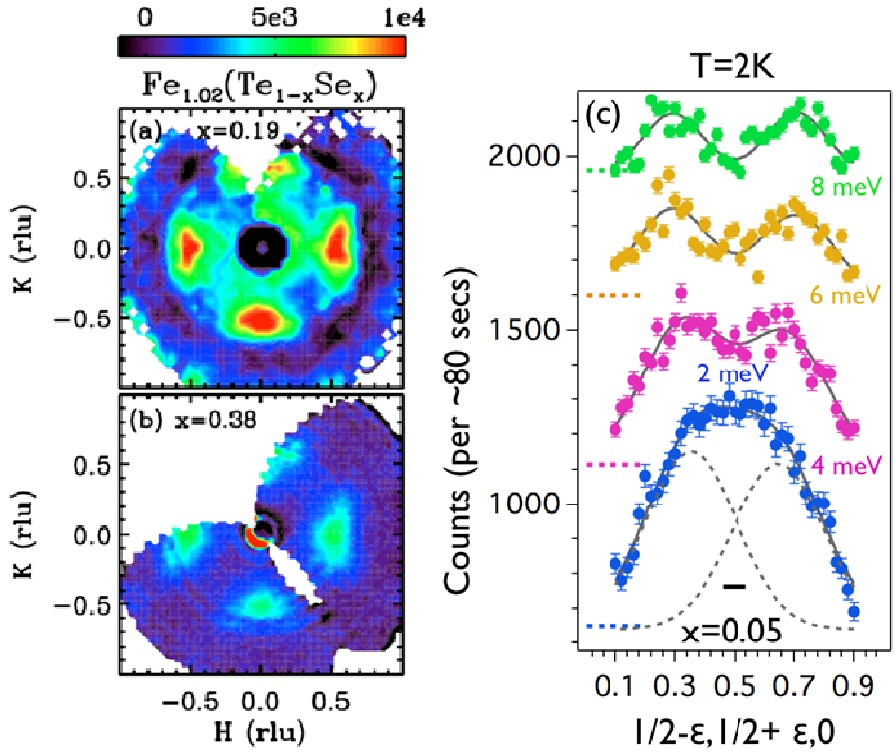

Figure 4: **Difference of microscopic magnetic properties between samples with and without bulk superconductivity**. **a**, Neutron scattering intensity map on the (HK0) plane for the $x=0.19$ sample without bulk superconductivity ($\hbar\omega= 0$ meV). **b**, Neutron scattering intensity map on the (HK0) plane for the $x=0.38$ sample with bulk superconductivity ($\hbar\omega = 0$ meV). Both samples exhibit quasi-elastic scattering near (1/2, 0), but the scattering in the $x=0.19$ sample is much stronger than that in the $x = 0.38$ sample. The data shown in **a** and **b** were measured using MACS spectrometer at the NIST. **c**. Typical INS transverse scans through (1/2, 1/2) (or ($\pi, \pi$)) at fixed energy measured from a single crystal of $Fe_{1.02}Te_{0.95}Se_{0.05}$ at 2K. The solid lines through the data are Gaussian fits to the magnetic excitations. For clarity, data and fits are shifted along the *y*-axis by an arbitrary amounts. The position of the background for each scan is indicated by a horizontal line. The black horizontal bar in panel represents the expected Q-resolution.